\documentclass[letter,twocolumn]{jpsj2} %% two-column layout
\newcommand{\im}{\textrm{i}}
\newcommand{\Ham}{\mathcal{H}}
\newcommand{\bs}[1]{\boldsymbol{#1}}
\newcommand{\bra}[1]{\big<#1\big|}
\newcommand{\ket}[1]{\big|#1\big>}
\newcommand{\hc}[1]{#1^\dagger}
\newcommand{\abs}[1]{\left|#1\right|}
\newcommand{\dsum}{\displaystyle\sum}
\newcommand{\up}{\uparrow}
\newcommand{\down}{\downarrow}
\newcommand{\nn}{\nonumber}

\title{Possibility of Gapless Spin Liquid State by One-dimensionalization}

\author{Yuta \textsc{Hayashi}\thanks{E-mail address: yhayashi@hosi.phys.s.u-tokyo.ac.jp} and Masao \textsc{Ogata}}

\inst{Department of Physics, University of Tokyo, Hongo, Bunkyo-ku, Tokyo, 113-0033}

\abst{Motivated by the observation of a gapless spin liquid state in $\kappa$-(BEDT-TTF)$_2$Cu$_2$(CN)$_3$, we analyze the anisotropic triangular lattice $S=1/2$ Heisenberg model with the resonating valence bond mean-field approximation. Paying attention to the small quasi-one-dimensional anisotropy of the material, we take an approach from one-dimensional (1D) chains coupled with frustrating zig-zag bonds. By calculating one-particle excitation spectra changing anisotropy parameter $J'/J$ from the decoupled 1D chains to the isotropic triangular lattice, we find almost gapless excitations in the wide range from the 1D limit. This one-dimensionalization by frustration is considered to be a candidate for the mechanism of the gapless spin liquid state.}

\kword{gapless spin liquid, $\kappa$-(BEDT-TTF)$_2$Cu$_2$(CN)$_3$, anisotropic triangular lattice, frustration, one-dimensionalization}

\begin{document}

\maketitle

Organic conductors are one of the fascinating materials which have low-dimensionality and relatively strong electron correlations. So far, various physical states have been observed and investigated intensively\cite{review}. Among them, magnetism in the Mott insulating phase next to the unconventional superconductivity has been attracting considerable attention. This phase is observed in the family of $\kappa$-(BEDT-TTF)$_2$X, where BEDT-TTF (ET) denotes bis(ethylenedithio)-tetrathiafulvalene and X represents a monovalent anion. Similarities to that of high-$T_{\textrm{c}}$ cuprates are worthy of note. Another stimulating problem concerning magnetism is ground state properties of geometrically frustrated spin systems such as a triangular lattice and a Kagom\'e lattice. These two intriguing issues meet in a material $\kappa$-(ET)$_2$Cu$_2$(CN)$_3$, which is a Mott insulator having a nearly isotropic triangular lattice, and it has been in the spotlight of late.

According to $^1$H NMR measurements at ambient pressure\cite{Cu2CN3_1}, $\kappa$-(ET)$_2$Cu$_2$(CN)$_3$ shows no indication of long-range magnetic order (LRMO) down to 32\,mK. This is 4 orders of magnitude below the exchange constant $J\sim250$\,K estimated from the temperature dependence of susceptibility. Recently, a similar result has been obtained by zero-field muon spin relaxation measurements, which have observed no LRMO down to 20\,mK\cite{muSR}. These results suggest that a quantum spin liquid state is realized in the ground state. On the other hand, the static susceptibility remains finite down to 1.9\,K, and spin-lattice relaxation rate $1/T_1$ shows power-law temperature dependence below 1\,K. These imply that almost gapless spin excitation exists. This fact is a significant feature of the spin liquid phase observed in this material.

Since Anderson's proposal of a resonating valence bond (RVB) state\cite{Anderson}, enormous number of studies have been made on the triangular lattice spin system. It is now a general view that the ground state of the isotropic triangular lattice Heisenberg model has LRMO, such as the 120$^\circ$ structure\cite{ED,SE,LSW,VMC1}. On the other hand, if one neglects the LRMO and assumes a disordered ground state, the mean-field theory of RVB state gives a spin-gap state with d$_{x^2-y^2}$+id$_{xy}$-wave symmetry, which is called ``d+id state''\cite{tJ_RVB1,tJ_RVB2,VMC2}. This RVB state, describing an insulating spin system, corresponds to a projected BCS state at half-filling in which doubly occupied states are excluded. Thus, the existing theories show that the ground state has LRMO in general, and if the magnetic order is destroyed in some reason, the d+id fullgap state will appear. If we regard the Mott insulating phase of $\kappa$-(ET)$_2$Cu$_2$(CN)$_3$ in low temperatures as an isotropic triangular lattice spin system, the results of NMR and susceptibility measurements, which suggest neither LRMO nor spin gap, cannot be explained.

\begin{table}[t]
	\begin{center}
		\caption{Anisotropy of effective transfer integrals in $\kappa$-(ET)$_2$X\cite{aniso}. The definition of $t$ and $t'$ are not as usual (see the text).}
		\vspace{2mm}
		\begin{tabular}{cccc}
			\hline
				Anion X					&&	$t'/t$	&	\\
			\hline
				Cu$_2$(CN)$_3$			&&	0.94	&	\\
				Cu(NCS)$_2$				&&	1.19	&	\\
				Cu[N(CN)$_2$]Br			&&	1.33	&	\\
				Cu[N(CN)$_2$]Cl			&&	1.47	&	\\
				Cu(CN)[N(CN)$_2$]		&&	1.47	&	\\
				Ag(CN)$_2\cdot$H$_2$O	&&	1.67	&	\\
				I$_3$					&&	1.72	&	\\
			\hline
		\end{tabular}
		\label{tt}
		\vspace{-1mm}
	\end{center}
\end{table}

In this letter, we pay attention to small anisotropy of $\kappa$-(ET)$_2$Cu$_2$(CN)$_3$ and propose a new possibility for understanding its gapless spin liquid state. As shown in Table \ref{tt}, only $\kappa$-(ET)$_2$Cu$_2$(CN)$_3$ has an opposite anisotropy among the family of $\kappa$-(ET)$_2$X studied in the past. Here, the effective transfer integrals $t$ and $t'$ are defined inversely to the conventional way; $t=0$ corresponds to the square lattice, and $t'=0$ the decoupled chains. Therefore, $\kappa$-(ET)$_2$Cu$_2$(CN)$_3$ has quasi-one-dimensional (Q1D) anisotropy rather than an isotropic triangular lattice.
Considering that the pure 1D spin system has no LRMO and gapless spin excitation, it is likely that this Q1D anisotropy is concerned with the formation of the gapless spin liquid state in $\kappa$-(ET)$_2$Cu$_2$(CN)$_3$.

\begin{figure}[b]
	\begin{center}
		\vspace{-1mm}
		\includegraphics[width=60mm]{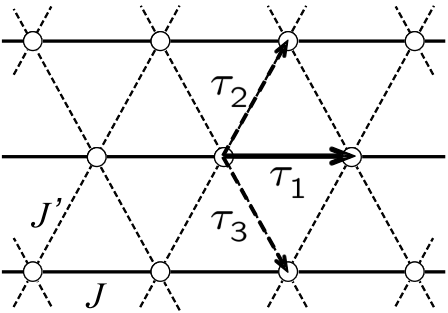}
		\vspace{-2mm}
		\caption{The anisotropic triangular lattice Heisenberg model with intrachain coupling $J$ and interchain zig-zag coupling $J'$. $\bs{\tau}_1,\,\bs{\tau}_2,\,\bs{\tau}_3$ are lattice vectors.}
		\label{fig1}
	\end{center}
\end{figure}

Based on the above consideration, we study the Heisenberg model on an anisotropic triangular lattice, which is equivalent to 1D chains coupled with zig-zag bonds as shown in Fig. \ref{fig1}. The Hamiltonian is given by
\begin{equation}
	\Ham = \!\!\sum_{<i,i'>}\!\!J\bs{S}_i\cdot\bs{S}_{i'}+\!\sum_{<i,j>}\!\!J'\bs{S}_i\cdot\bs{S}_j,  \label{Ham}
\end{equation}
where $<\!\!i,i'\!\!>$ and $<\!\!i,j\!\!>$ represent the summation over intrachain and interchain nearest-neighbor pairs with antiferromagnetic coupling constant $J$ and $J'$, respectively (see Fig. \ref{fig1}). We investigate the anisotropy parameter range $J'/J$ = 0.0-1.0, in which the model interpolates between the decoupled chains ($J'=0$) and the isotropic triangular lattice ($J'=J$).

In the following, we consider a projected BCS state defined as
\begin{equation}
	\ket{\textrm{p-BCS}} \equiv P_{\textrm{G}}\ket{\textrm{BCS}},  \label{pBCS}
\end{equation}
where $P_{\textrm{G}}$ is the Gutzwiller projection operator which excludes double occupancy and $\ket{\textrm{BCS}}$ is a BCS mean-field wave function. Since it is difficult to treat the Gutzwiller projection analytically, we apply an RVB mean-field approximation to the Hamiltonian (\ref{Ham}) and calculate the one-particle excitation spectra. To put it more concretely, we introduce mean fields $\Delta_{ij} \equiv \big<c_{i\up}c_{j\down}\big>\,,\,\xi_{ij} \equiv \big<\hc{c_{i\up}}c_{j\up}\big> = \big<\hc{c_{i\down}}c_{j\down}\big>$ and obtain its excitation spectrum by diagonalizing the mean-field Hamiltonian. This approximation is equivalent to the ``Gutzwiller approximation'' which replaces the effect of the Gutzwiller projection operator with the statistical weight $g_{\textrm{s}}$ as
\begin{equation}
	\bra{\textrm{p-BCS}}\bs{S}_i\!\cdot\!\bs{S}_j\ket{\textrm{p-BCS}}
	= g_{\textrm{s}}\bra{\textrm{BCS}}\bs{S}_i\!\cdot\!\bs{S}_j\ket{\textrm{BCS}}.
\end{equation}
In the simplest Gutzwiller approximation, the statistical weight is given as $g_{\textrm{s}}=4/(1+\delta)^2$ where $\delta$ is the density of holes\cite{GA}, and in the case of half-filling ($\delta=0$), $g_{\textrm{s}}=4$. Although double occupancy is no longer excluded from wave functions in this approximation, it is known in the research of high-$T_{\textrm{c}}$ superconductivity that the RVB mean-field (Gutzwiller) approximation gives qualitatively good results.

The spin operators $\bs{S}_i\cdot\bs{S}_j$ in the Hamiltonian (\ref{Ham}) can be rewritten by the fermion operators as
\begin{align}
	\bs{S}_i\cdot\bs{S}_j
	&= \dfrac{1}{4}\!\left(\hc{c_{i\up}}c_{i\up}-\hc{c_{i\down}}c_{i\down}\right)\!\!
					\left(\hc{c_{j\up}}c_{j\up}-\hc{c_{j\down}}c_{j\down}\right)  \nn \\
	&\qquad	+\dfrac{1}{2}\!\left(\hc{c_{i\up}}c_{i\down}\hc{c_{j\down}}c_{j\up}
					+\hc{c_{i\down}}c_{i\up}\hc{c_{j\up}}c_{j\down}\right).
\end{align}
By introducing the mean fields, we can rewrite the Hamiltonian as
\begin{equation}
	\Ham_{\textrm{MF}}
	= \!\sum_{\bs{k}}\!\bigg[\xi_{\bs{k}}\!\left(\hc{c_{\bs{k}\up}}c_{\bs{k}\up}\!+\hc{c_{\bs{k}\down}}c_{\bs{k}\down}\right)\!
		+ \!\left(\Delta_{\bs{k}}\hc{c_{\bs{k}\up}}\hc{c_{-\bs{k}\down}}\!+\textrm{h.c.}\right)\!\!\bigg]
\end{equation}
except for constant terms. Here, $\xi_{\bs{k}}$ and $\Delta_{\bs{k}}$ are given by
\begin{align}
	\xi_{\bs{k}}
	&\equiv -3J\xi_{\bs{\tau}_1}\!\cos(\bs{k}\cdot\bs{\tau}_1)  \nn \\
	&\quad\,\,  -3J'\Big[\xi_{\bs{\tau}_2}\!\cos(\bs{k}\cdot\bs{\tau}_2)+\xi_{\bs{\tau}_3}\!\cos(\bs{k}\cdot\bs{\tau}_3)\Big],  \\
	\Delta_{\bs{k}}
	&\equiv 3J\Delta_{\bs{\tau}_1}\!\cos(\bs{k}\cdot\bs{\tau}_1)  \nn \\
	&\quad\,\,  +3J'\Big[\Delta_{\bs{\tau}_2}\!\cos(\bs{k}\cdot\bs{\tau}_2)+\Delta_{\bs{\tau}_3}\!\cos(\bs{k}\cdot\bs{\tau}_3)\Big],
\end{align}
where $\bs{\tau}_1=(1,0)$, $\bs{\tau}_2=(1/2,\sqrt{3}/2)$, $\bs{\tau}_3=(1/2,-\sqrt{3}/2)$ as shown in Fig. \ref{fig1}, and
\begin{equation}
	\xi_{\tau} \equiv \left<\hc{c_{i\up}}c_{i+\bs{\tau}\up}\right>
				\!=\! \left<\hc{c_{i\down}}c_{i+\bs{\tau}\down}\right>, \,\,\,
	\Delta_{\tau} \equiv \Big<c_{i\up}c_{i+\bs{\tau}\down}\Big>.
\end{equation}
On the analogy of BCS theory, we obtain self-consistent equations at zero temperature
\begin{equation}
	\left\{
	\begin{array}{l}
	\, \xi_{\bs{\tau}_i}
	= -\dfrac{1}{2N}\dsum_{\bs{k}}e^{\im\bs{k}\cdot\bs{\tau}_i}\dfrac{\xi_{\bs{k}}}{E_{\bs{k}}}  \\ [12pt]
	\Delta_{\bs{\tau}_i} \!
	= \dfrac{1}{2N}\dsum_{\bs{k}}e^{-\im\bs{k}\cdot\bs{\tau}_i}\dfrac{\Delta_{\bs{k}}}{E_{\bs{k}}},
	\end{array}
	\right.  \label{SCeqs}
\end{equation}
with a quasiparticle excitation spectrum
\begin{equation}
	E_{\bs{k}} = \sqrt{\xi_{\bs{k}}^2+\abs{\Delta_{\bs{k}}}^2}.  \label{Ek}
\end{equation}
We determine the order parameters $\Delta_{\bs{\tau}_i}\,,\,\xi_{\bs{\tau}_i}$ ($i=1,2,3$) by solving self-consistent equations (\ref{SCeqs}) numerically, and obtain the one-particle excitation spectrum $E_{\bs{k}}$. 

Firstly, we verify our method in 1D limit ($J'/J=0$). According to the exact solution, the ground state is a spin disordered state and the excitation spectrum is ``des Cloizeaux-Pearson mode'' with $S=1$\cite{dCP}. In the present RVB mean-field theory, the one-particle excitation spectrum becomes
\begin{equation}
	E_{\bs{k}} = 3J\sqrt{\xi_{\bs{\tau}_1}^2+\abs{\Delta_{\bs{\tau}_1}}^2}\abs{\cos k_x}
\end{equation}
in the 1D limit. This clearly realizes gapless excitations at $k_x=\pm\pi/2$. Note that this one-particle excitation describes a spin singlet breaking, i.e. \,$S=1/2$ spinon excitation, whereas the des Cloizeaux-Pearson mode describes $S=1$ spin-wave (magnon) excitation. Thus, two-spinon excitations with $k_x=\pi/2$ and $k_x=-\pi/2$ form an $S=1$ magnon with $k_x=0$. This means that the present gapless excitation spectrum obtained in the RVB mean-field theory is consistent with the exact des Cloizeaux-Pearson mode.

\begin{figure}[b]
	\begin{center}
		\vspace{-1mm}
		\includegraphics[width=60mm]{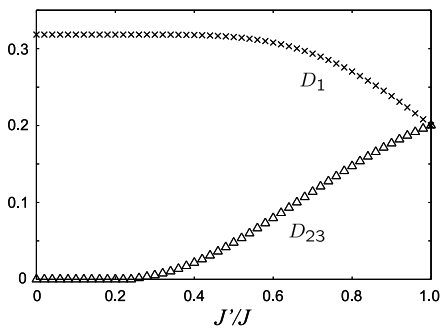}
		\vspace{-3mm}
		\caption{Anisotropy dependence of $D_1$ and $D_{23}$ for $L=1200$. Note that $D_{23}$ is very small compared to $D_1$ in a wide range $0\leq J'/J\lesssim$0.25.}
		\label{fig2}
	\end{center}
\end{figure}

Nextly, we show the results of $0\leq J'/J\leq1$ case, focusing on the following parameters
\begin{equation}
	\left\{
	\begin{array}{l}
		D_1 \,\equiv \sqrt{\xi_{\bs{\tau}_1}^2+\abs{\Delta_{\bs{\tau}_1}}^2}  \nn \\ [5pt]
		\! D_{23} \equiv \sqrt{\xi_{\bs{\tau}_2}^2+\abs{\Delta_{\bs{\tau}_2}}^2}
						=\sqrt{\xi_{\bs{\tau}_3}^2+\abs{\Delta_{\bs{\tau}_3}}^2}.
	\end{array}
	\right.  \label{D}
\end{equation}
Because of the SU(2) degeneracy at half-filling\cite{tJ_RVB1,GA}, these parameters are determined uniquely regardless of the degenerate ground states. Actually, the excitation spectrum can be written as
\begin{align}
	E_{\bs{k}}^2
	&= 9J^2D_1^2\cos^2\!k_x  \nn \\
	&\, +9J'\/^2D_{23}^2\!\left[
		\cos^2\!\!\left(\!\dfrac{k_x}{2}\!+\!\dfrac{\sqrt{3}k_y}{2}\!\right)\!\!
		+\cos^2\!\!\left(\!\dfrac{k_x}{2}\!-\!\dfrac{\sqrt{3}k_y}{2}\!\right)\!\right].  \label{ED123}
\end{align}
Therefore, $D_1$, $D_{23}$ determine the dispersion relations along the chains ($\bs{\tau}_1$) and between the chains ($\bs{\tau}_2$,$\bs{\tau}_3$), respectively. Their $J'/J$ dependence calculated in the system size $L=1200$ ($N=L^2$) are plotted in Fig. \ref{fig2}. A notable feature is that $D_{23}$ remains very small compared to $D_1$, in spite of the comparatively large $J'$ up to $J'/J\sim0.25$. When $D_{23}=0$ the system is a pure 1D chain. Indeed, when $J'/J=0$, the right-hand side of the self-consistent equations of $\xi_{\bs{\tau}_2}$, $\xi_{\bs{\tau}_3}$, $\Delta_{\bs{\tau}_2}$, $\Delta_{\bs{\tau}_3}$ become all equal to zero. As we show later, $D_{23}$ is very small for $J'/J\lesssim0.25$ and vanishes when $J'/J\to0$. This indicates that there are scarcely any correlations between spins of different chains, and practically 1D state is realized. As $J'/J$ approaches unity, $D_{23}$ gradually increases and becomes equal to $D_1$.

\begin{figure}[b]
	\begin{center}
		\includegraphics[width=80mm]{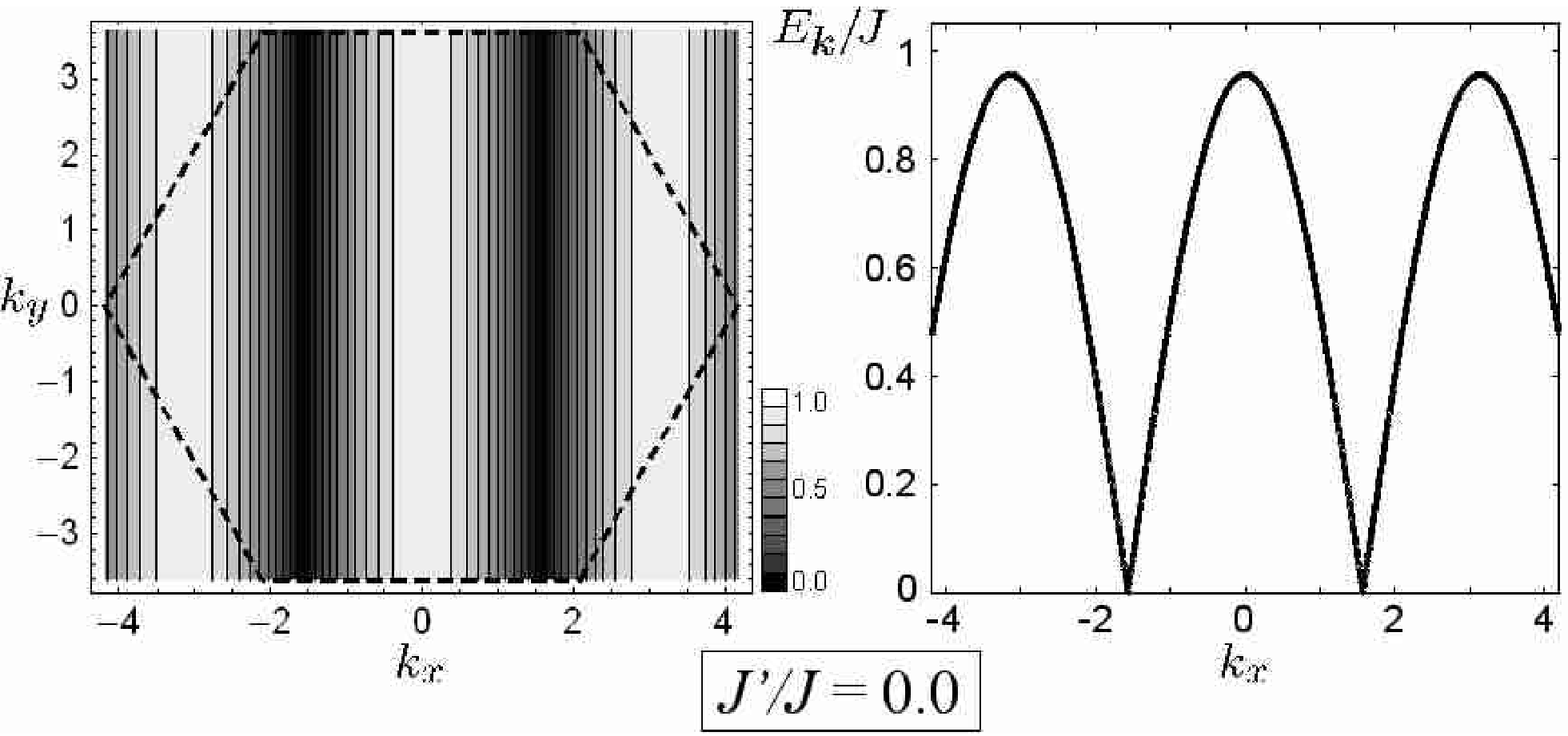} \\\vspace{2mm}
		\includegraphics[width=80mm]{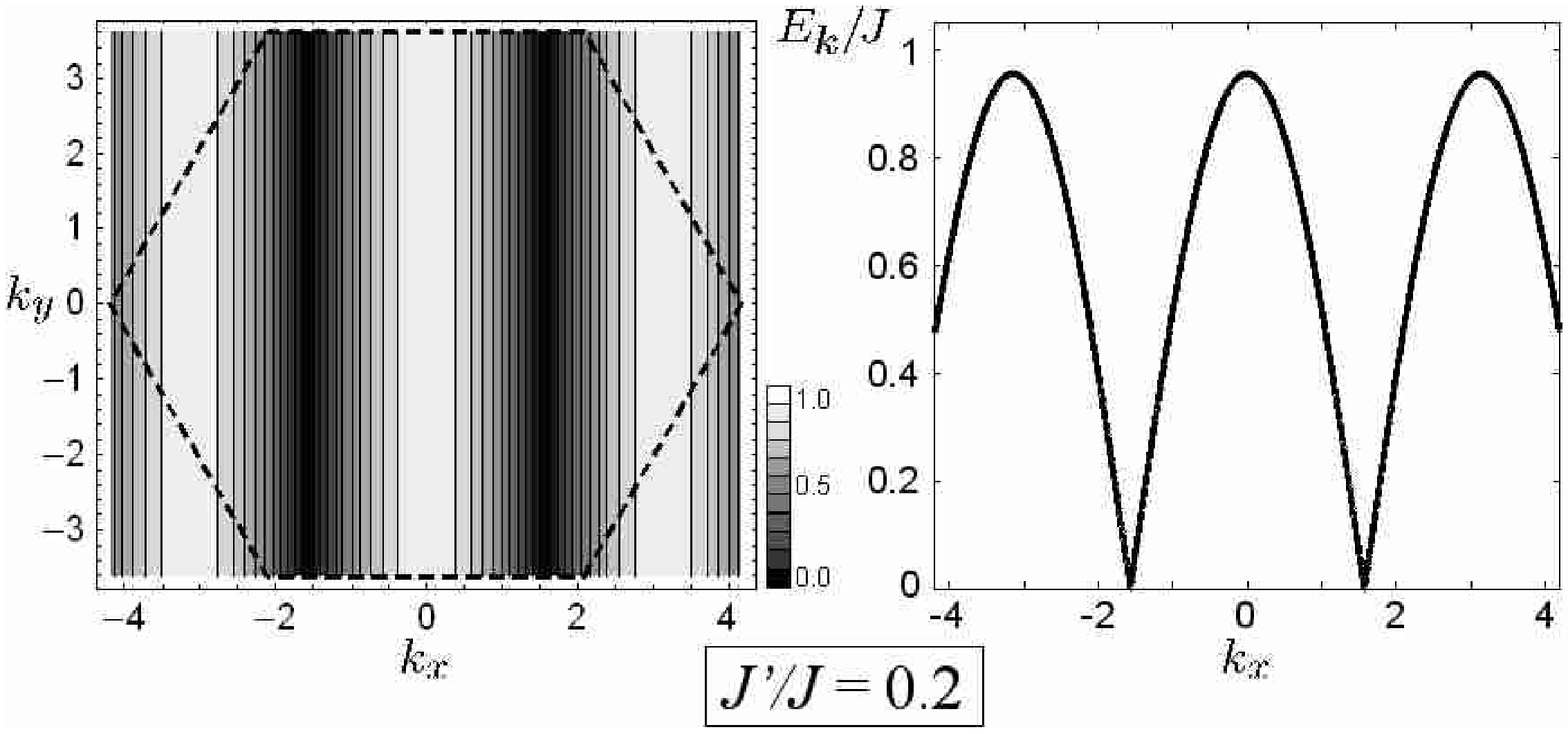} \\\vspace{2mm}
		\includegraphics[width=80mm]{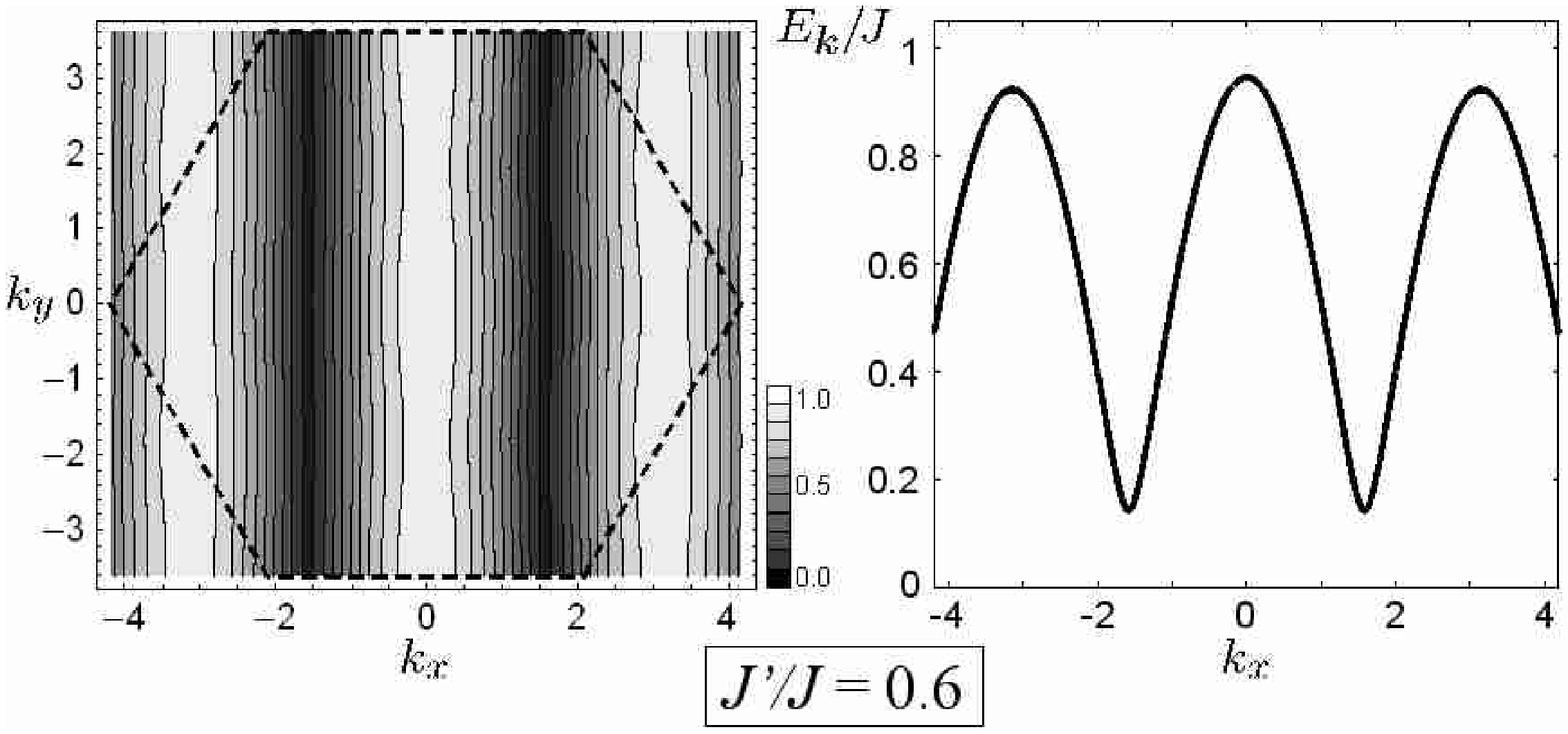} \\\vspace{2mm}
		\includegraphics[width=80mm]{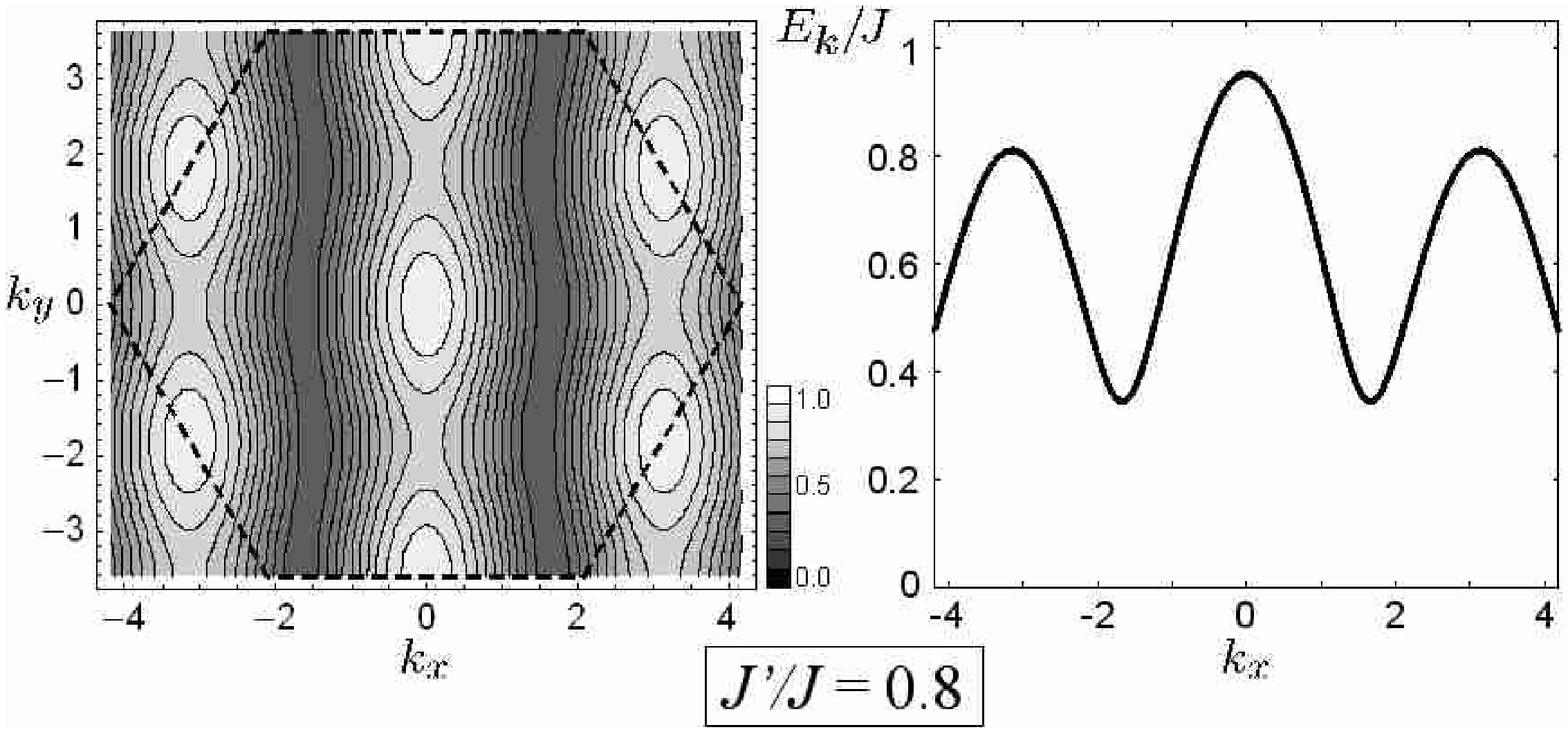} \\\vspace{2mm}
		\includegraphics[width=80mm]{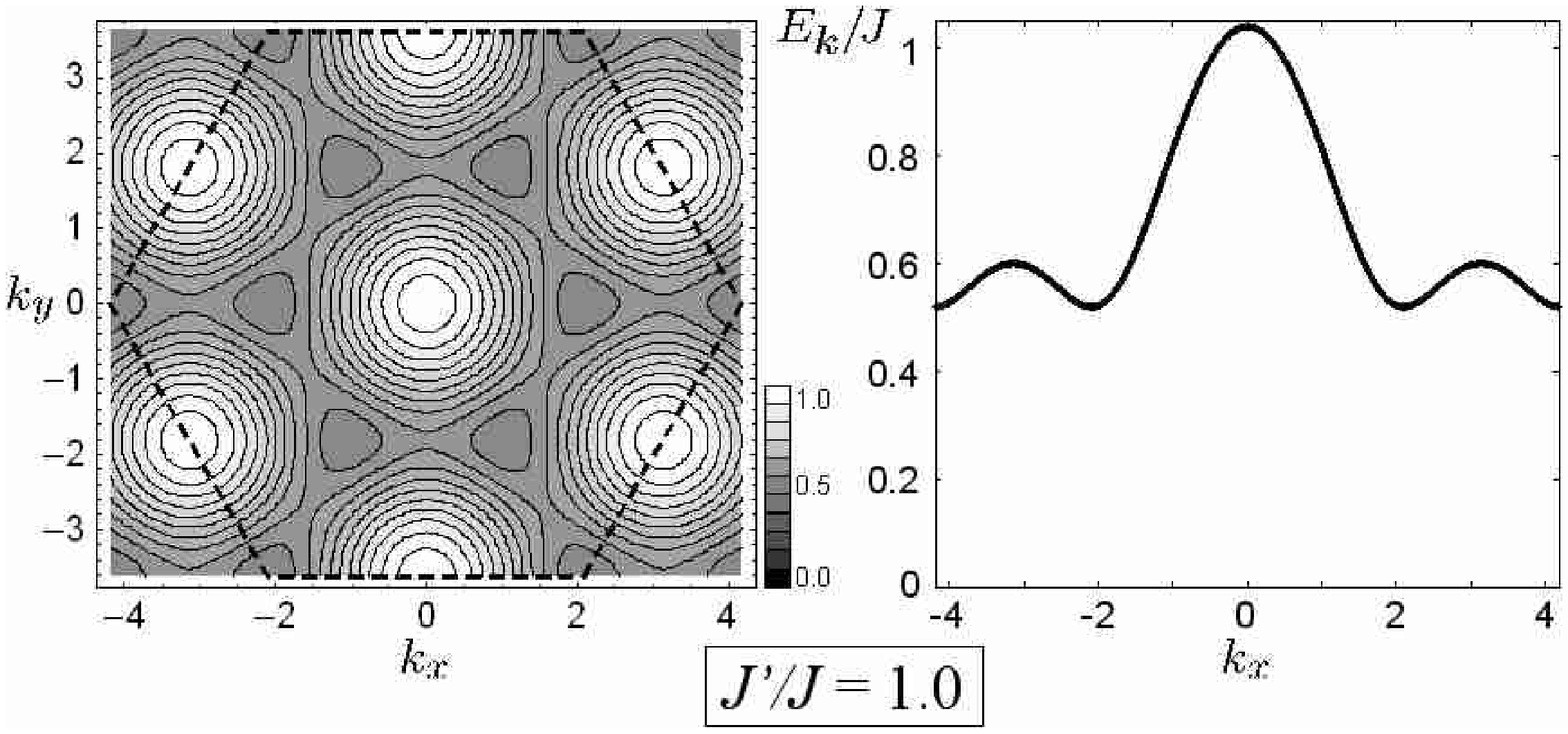}
		\vspace{-1mm}
		\caption{Anisotropy dependence of the one-particle excitation spectra. Contour plots of the spectra are on the left, and sections along $k_y=0$ line are on the right. The hexagons with broken lines represent 1BZ of the triangular lattice. Up to $J'/J\sim0.25$, the spectra for each anisotropy are hardly distinguishable, and the one-dimensionality strongly remains for large $J'/J$. }
		\label{fig3}
	\end{center}
\end{figure}

Finally, we show in Fig. \ref{fig3} the $J'/J$ dependence of the one-particle excitation spectra $E_{\bs{k}}$ in (\ref{ED123}). We find that the structure of excitation spectra in $0\leq J'/J\lesssim0.25$ has little difference from that of the decoupled chains ($J'/J=0.0$). As a result, almost gapless excitations are realized in this wide parameter range. This means that practically 1D state is realized, which is also expected from the behavior of $D_{23}$ in Fig. \ref{fig2}. When $J'/J$ exceeds 0.25, the excitation gap gradually increases globally in the first Brillouin zone (1BZ). However, the shape of the whole spectrum is almost unchanged until the $J'/J$ becomes as large as about 0.6. Moreover, focusing on the lowest energy excitations (dark areas in the contour plot shown in Fig. \ref{fig3}), their locations in the 1BZ do not deviate from those in the 1D limit ($k_x=\pm\pi/2$) for $J'/J\lesssim0.8$. Additionally, when $k_x=\pm\pi/2$, the excitation spectrum $E_{\bs{k}}$ is independent of $k_y$, i.e., $E_{\bs{k}}=3J'D_{23}$. This is because the frustration of two interchain couplings (corresponding to the lattice vector $\bs{\tau}_2$ and $\bs{\tau}_3$) cancel the $k_y$ dependence. This fact is rather important, since it indicates that the excited quasiparticles along the $k_x=\pm\pi/2$ lines feel free to move along the $k_y$ direction. This is the same condition as in the 1D limit, except for the existence of a finite energy gap.

\begin{figure}[b]
	\begin{center}
		\vspace{-1mm}
		\includegraphics[width=75mm]{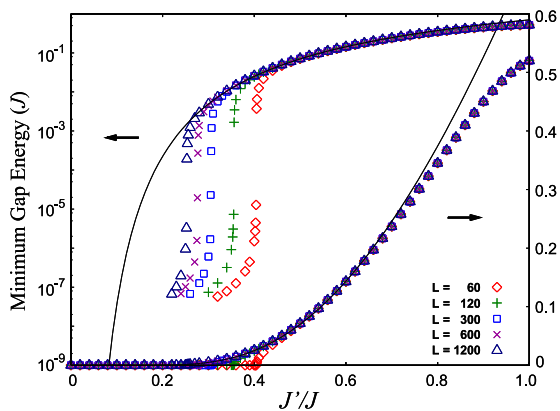}
		\vspace{-3mm}
		\caption{(Color Online) Anisotropy dependence of the minimum gap energy in the 1BZ (right axis) for $L$=60(diamond), 120(plus), 300(square), 600(cross) and 1200(triangle). The semi-log plots of the same quantity are also shown (left axis). The solid line is a fitted exponential function $aJ'\exp(-bJ/J')$, where $a=3.50$ and $b=1.61$. We find that the observed critical behavior is an artifact of finite size calculation (see the text).}
		\label{fig4}
	\end{center}
\end{figure}

Figure \ref{fig4} shows the minimum gap energy in the 1BZ as a function of anisotropy $J'/J$, changing the system size $L$. We can see the almost gapless excitations in the wide parameter range $0\leq J'/J \lesssim0.25$, as is already expected. It is quite natural that this behavior is similar to that of $D_{23}$, considering that the minimum energy excitations are located along $k_x=\pm\pi/2$ for $J'/J\lesssim 0.6$. By plotting the same data for various system size, $L$, in a semi-log scale (Fig. \ref{fig4}), we can see a discontinuous jump for every size. We find that this critical value $J'_c/J$ vanishes very slowly as $(\ln L)^{-1}$. Thus, the discontinuity is an artifact of finite-size calculation.
We also find that the minimum gap energy is finite when infinitesimal $J'$ is introduced. Actually, we can fit the $J'$ dependence as $aJ'\exp(-bJ/J')$\cite{Misawa} for $J'/J\lesssim 0.6$ as shown in Fig. \ref{fig4}.
Considering that the minimum gap energy is already about 3 orders of magnitude below $J$ at $J'/J\sim0.25$, it can be said that almost gapless excitation is realized in $0\leq J'/J \lesssim0.25$. This result is fairly suggestive compared with the previous series expansion\cite{J1J2_SE} and linear spin wave\cite{J1J2_LSW1,J1J2_LSW2} studies, all of which suggest a spin disordered state in the parameter range $J'/J\lesssim0.25$.

From the above results, we conclude that there is a strong tendency to form a 1D-like excitation spectrum for the triangular lattice spin system with anisotropy $0\leq J'/J\lesssim 0.6$. Furthermore, even if the anisotropy is as large as $0.6\lesssim J'/J \lesssim 0.8$, we can still expect 1D-like behavior for quasiparticles except for the existence of the excitation gap.
Let us here discuss the relation to $\kappa$-(ET)$_2$Cu$_2$(CN)$_3$. The anisotropy of spin exchange interactions in this material can be estimated from $J=4t^2/U$ ($U$ being the onsite Coulomb repulsion) as $J'/J\sim0.89$. At this anisotropy, a rather large excitation gap exists as shown in Fig. \ref{fig4}.
We consider two possibilities to understand the gaplessness. One is that the small gap region in Fig. \ref{fig4} expands to large values of $J'/J$ by some factors not considered in the present model. For example, If long-distance exchange interactions, quantum fluctuation or multiple spin exchange effect\cite{MSE} (higher order terms of the Heisenberg model) suppress not only LRMO but also the spin gap, we can reproduce the gapless spin liquid state at large $J'/J$. These possibilities remain as future problems. Another possibility is that the anisotropy $J'/J$ of $\kappa$-(ET)$_2$Cu$_2$(CN)$_3$ deviates from the above estimation due to, for example, a finite $U$ effect\cite{t^3}. If it is in the range $J'/J<0.25$, the excitation gap is sufficiently small and the susceptibility behavior (finite at 1.9\,K whereas $J\sim250$\,K) can be explained.

In summary, we analyzed an anisotropic triangular lattice Heisenberg model using RVB mean-field approximation in order to investigate the physical origin of the gapless spin liquid state observed in $\kappa$-(ET)$_2$Cu$_2$(CN)$_3$. We payed attention to the Q1D anisotropy of this material, and took an approach from the 1D limit. As a result of calculations, we found that a practically 1D state with almost gapless excitations is realized in the wide range of the anisotropy parameter $0\leq J'/J \lesssim0.25$. Furthermore, one-dimensionality remained strongly even in $J'/J>0.25$ due to the geometrical frustration of interchain couplings. We consider this ``one-dimensionalization by frustration'' as a candidate for the mechanism of the gapless spin liquid state, although the full understanding has not yet been achieved.

This work was partly supported by a Grant-in-Aid for Scientific Research on Priority Areas of Molecular Conductors (No. 15073210) from the Ministry of Education, Culture, Sports, Science and Technology, Japan, and also by a Next Generation Supercomputing Project, Nanoscience Program, MEXT, Japan.


\begin{thebibliography}{99}
\bibitem{Misawa} We would like to thank T.Misawa for pointing out this possibility.
\end{thebibliography}
\end{document}